\documentclass[twocolumn]{aastex62}

\graphicspath{{./}{figures/}}

\shorttitle{Constraints from NIR fast photometry of GX 339--4}
\shortauthors{Vincentelli et al.}
\begin{document}

\title{Physical constraints from near-infrared fast photometry of the black-hole transient GX 339--4}

\correspondingauthor{Federico M. Vincentelli}
\email{F.M.Vincentelli@soton.ac.uk}

\author[0000-0002-1481-1870]{F. M. Vincentelli}
\affiliation{Department of Physics and Astronomy, University of Southampton, Southampton SO17 1BJ}
%\collaboration{(AAS Journals Data Scientists collaboration)}

\author[0000-0002-0752-3301]{P. Casella}
\affiliation{INAF, Osservatorio Astronomico di Roma 
Via Frascati 33, I-00078 Monteporzio Catone, Italy}
%\nocollaboration

\author[0000-0001-6061-3480]{P. Petrucci}
\affiliation{Univ. Grenoble Alpes, CNRS, IPAG, F-38000 Grenoble, France.}

\author[0000-0003-0976-4755]{T. Maccarone}
\affiliation{Texas Tech University, Physics \& Astronomy Department, Box 41051, Lubbock, TX 79409-1051}

\author[0000-0002-3500-631X]{D. M. Russell}
\affiliation{Center for Astro, Particle and Planetary Physics, New York University Abu Dhabi, PO Box 129188, Abu Dhabi, UAE}

\author{P. Uttley}
\affiliation{Astronomical Institute, Anton Pannekoek, University of Amsterdam, Science Park 904, NL-1098 XH Amsterdam, Netherlands}

\author[0000-0003-2743-6632]{B. De Marco}
\affiliation{Nicolaus Copernicus Astronomical Center, Polish Academy of Sciences, Bartycka 18, PL-00-716 Warsaw, Poland}

\author{R. Fender}
\affiliation{Department of Physics, Astrophysics, University of Oxford, Denys Wilkinson Building, Keble Road, Oxford, OX1 3RH, UK}

\author[0000-0003-3105-2615]{P. Gandhi}
\affiliation{Department of Physics and Astronomy, University of Southampton, Southampton SO17 1BJ}

\author[0000-0003-3334-3424]{J. Malzac}
\affiliation{IRAP, Université de Toulouse, CNRS, UPS, CNES, Toulouse, France}

\author[0000-0002-4744-3429]{K. O'Brien}
\affiliation{Department of Physics, Durham University, South Road, Durham, DH1 3LE, UK}

\author[0000-0001-5506-9855]{J. A. Tomsick}
\affiliation{Space Sciences Laboratory, 7 Gauss Way, University of California, Berkeley, CA 94720, USA}

%\nocollaboration



\begin{abstract}
We present results from the first multi-epoch X-ray/IR fast-photometry campaign on  the black-hole transient GX 339--4, during its 2015 outburst decay. We studied the evolution of the power spectral densities finding strong differences between the two bands. The X-ray power spectral density follows standard patterns of evolution, plausibly reflecting changes in the accretion flow. The IR power spectral density instead evolves very slowly, with a high-frequency break consistent with remaining constant at  $0.63 \pm 0.03$ Hz throughout the campaign. We discuss this result in the context of the currently available models for the IR emission in black-hole transients. {While all models will need to be tested quantitatively against this unexpected constraint, we show that an IR emitting relativistic jet which filters out the short-timescales fluctuations injected from the accretion inflow appears as the most plausible scenario}.

%and argue that the most plausible explanation is in terms of a nearly constant jet-launching radius, which we estimate being of the order of $10\,R_G$.
\end{abstract}

\keywords{accretion, accretion disks --- black hole physics --- stars: individual (GX 339--4) --- X-rays: binaries }

\section{Introduction}% \label{sec:intro}

Black-hole X-ray binaries (BHXRBs), are systems in which a stellar-mass black-hole has a companion star in close orbit, leading to the transfer of mass from the star to the hole. They are strong multi-wavelength emitters, showing a complex and variable spectrum from radio frequencies to hard X-rays \citep[see e.g.][and references therein]{fender2000,markoff2001,corbel2002,gandhi2011,corbel2013}. The time-averaged spectrum is the result of variable broad-band emission from a number of physical components, each emitting over broad and overlapping energy ranges. In their so-called ``hard states'', the X-ray spectra of BHXRBs are dominated by a power-law component, with a cutoff at around 100 keV\citep{motta2009,kalemci2014}, which is generally believed to arise from an optically thin, geometrically thick inflow \citep{thorne1975,narayan1995,zdiarski2004,dgk07}. This hot %, possibly magnetized 
plasma Comptonizes thermal, colder (ultraviolet to soft X-rays) photons from an accretion disk and perhaps also less energetic (optical or even infrared) synchrotron photons from the inflow itself \citep{malzac2009,poutanen2009,veledina2011,veledina2013}. At longer wavelengths, a flat (or slightly inverted) spectrum from radio down to optical-to-infrared (O-IR) can be seen, representing clear evidence for the presence of a compact jet \citep{corbel2002,gandhi2011,russellt2013}.
%, potentially contributing also at shorter wavelengths \citep{markoff2001,corbel2002,gandhi2011,russellt2013}
Consensus has not been reached yet on the quantitative contribution of each of these spectral components at different wavelengths, nor on their evolution.

%Jets from black-hole X-ray binaries (BHXRBs) attracted a wealth of scientific attention since their discovery, due to their rich and complex phenomenology, especially their transient nature - which permits to study their behaviour at different regimes. BHXRBs emit most of their radiation at X-rays wavelengths, due to the high temperatures reached by matter accreted near to the black-hole. They display during their outbursts two main spectral states, which are also strongly related to the manifestation of jets \citep{fender2004}. In the hard state, the X-ray spectrum is dominated by a power-law component which extends up to 100 keV and is generally believed to be due to a optically thin, geometrically thick inflow which comptonises the thermal photons from an outer accretion disk. At longer wavelengths, a flat (or slightly inverted) spectrum from radio down to O-IR can be seen, clear evidence of a steady compact jet. In the soft state, the X-ray spectrum is instead dominated by the thermal emission of a standard Shakura-Sunayev accretion disk \citep{tanaka,homan2001}, and no radio emission is usually observed, a probable signature of a quenched jet \citep{Tananbaum1972}.  

Studies of the fast variability properties of the X-ray emission have been historically very important - albeit inconclusive - to constrain physical models \citep{nowak1999,churazov2001,ingram2009}.
The X-ray Fourier power spectral densities (PSD) during the hard state reveal a combination of broad-band components and narrow Quasi Periodic Oscillations (QPOs).
%which, modelled with a number of Lorentzian components, indicate a number of characteristic timescales. 
The behaviour of these components is closely related to the spectral evolution of these sources, with most of their characteristic frequencies increasing with the accretion rate and/or with the softening of the X-ray emission \citep{belloni2002,belloni2005}. The characteristic frequencies of the X-ray broad-band noise are often associated with the viscous timescales at some radii in the accretion flow, so that their evolution is explained in terms of a geometrical evolution of the accretion flow, which is thought to become more and more compact while the source evolves from a hard to soft spectral state. This interpretation is supported by the fact that the highest of these frequencies (the so-called "high-frequency break" in the X-ray PSD) appears to saturate at a few Hz \citep{churazov2001,belloni2005,ingram2012}: a value consistent with the viscous timescale for a hot optically thin, geometrically thick inflow, at its innermost stable orbit around a stellar-mass black-hole \citep{dgk07}. Further refinements of this broad picture came from more advanced X-ray spectral-timing techniques, where delays are studied as a function of both Fourier frequency and energy \citep[see e.g.][and references therein]{nowak1999,kotov2001,uttley2014,demarco2017,kara2019,Mahmoud2019}.

An alternative interpretation for the broad-band components observed in the X-ray PSD has been recently suggested by \cite{veledina2016} in terms of interference between two Comptonization continua. Both these components would respond to the fluctuations of the mass accretion rate, but with a variable delay between them due to the evolving propagation timescale from the disk Comptonization radius to the synchrotron Comptonization region.

Strong rapid variability has been also observed at longer, O-IR wavelengths \citep{motch1982,gandhi2008,casella2010,gandhi2016}. 
The observed phenomenology is rather complex, with almost all the above-mentioned spectral components potentially contributing to the measured variability: thermally reprocessed emission from the outer disk, and synchrotron emission from the jet and from a magnetized inflow. In some cases, an inflow could reproduce neatly the data \citep{veledina2017}, while in other cases a jet origin has been securely identified, at least for the fastest (sub-second) variability component \citep{gandhi2008} and especially at infrared wavelengths \citep{casella2010}, often showing a $\sim$\,0.1-second delay with respect to the X-ray variability. In at least some cases, an internal-shock jet model could reproduce successfully both the short timescale variability and the average broad-band spectral energy distribution \citep{malzac2014,malzac2018}. Further datasets have yielded further constraints on the geometry of relativistic jets \citep{gandhi2017}, and on the physical processes taking place at their base \citep{vincentelli2018}.

Altogether, the study of the multi-wavelength fast variability from BHXRBs has developed rapidly over the last years, showing a large potential for studying the physics of the accretion-ejection processes. Most of the results mentioned above were obtained with single observations, done on different sources or during different outbursts, which hampered investigations on the evolution of the various components. In this work we present results from a 1-month long monitoring campaign of the X-ray and IR fast variability properties of GX 339--4 during the decay of its 2015 outburst. During this phase, BHXRBs show a re-brightening in radio followed by an increase in luminosity at O-IR wavelength \citep{corbel2013}.
%The O-IR emission can be interpreted as synchrotron radiation from either the base of the jet \citep{coriat2009,russell2011,russell_2012} or a hot-magnetized inflow \citep{veledina2013,poutanen2014}.

%The structure of the letter is the following: follows: after a description of the data set,
%in Section 3.1

%\begin{enumerate}
%\item improved citations for third party data repositories and software,

%\end{enumerate}

\section{Observations}

The campaign consisted of five quasi-simultaneous X-ray and IR observations spanning between 2015-09-02 (MJD 57267) and 2015-10-03 (MJD 57298). The X-ray observations were carried out with XMM-Newton
%and Nustar
(P.I. Petrucci), while the IR data were collected with HAWK-I@VLT (Program ID 095.D-0211(A), P.I. Casella). The full X-ray dataset has been already described in detail by \cite{demarco2017}, who analyzed the evolution of the reverberation lag,  by \cite{Stiele2017}, who reported timing and spectral analysis, and by \cite{Wang-Ji2018}, who focused on the spectral properties. For an overview of the whole X-ray dataset we refer the reader to these works. Here we focus on the properties of the IR variability and its comparison with the X-ray variability.
%, thus we excluded from our analysis the first X-ray epoch, which was not contemporaneous to any HAWK-I epoch.
%Also, we limit our analysis to the XMM-Newton data, as Nustar data do not add additional relevant information in the context of the present work.

\subsection{X-ray Observations}

We extracted data from the XMM-Newton Epic-pn camera \citep{struder2001}.  The first three observations (2nd, 7th and 12th of September 2015) were taken in \textit{Timing} mode, while the last two (17th and 30th of September 2015) in \textit{Small window mode}. Following the procedure described in \cite{demarco2017}, we extracted the events in the 2-10 keV band, in a box of $\approx 86$ arcseconds of angular size in the first case (RAWX between 28 and 48), and within a circle of 40 arcseconds around the source in the latter. The extracted curves were barycentred to the Dynamical Barycentric Time system, using the command \textit{barycen}. All X-ray light curves were extracted with the highest time resolution available: i.e. 5.7 ms for the first three observations and 7.8 ms for the last two.

\subsection{IR data}

We collected IR  ($K_s$ band) high time resolution data with HAWK-I mounted at VLT UT-4/Yepun. HAWK-I is a near-infrared wide-field imager (0.97 to 2.31 $\mu$m) made by four HAWAII 2RG 2048x2048 pixel detectors \citep{Pirard2004}. In order to reach sub-second time resolution, the observations were performed  in \textit{Fast-Phot} mode, reading only a stripe made of 16 contiguous windows of 128 x 64 pixels in each quadrant. This allowed us to reach a time resolution as short as 0.105 seconds in the first two epochs (6th and 7th of September), 0.125 seconds for the 3rd and 4th (17th and 22nd of September), and 0.25 seconds for the last epoch (3rd of October). The data have a short ($\approx$3-second long) gap every 250 exposures, to empty the instrument buffer. {Gaps where then filled using the method described in \citep{kalamkar2016}.}
The instrument was pointed as to place 
%to the following coordinates (RA DEC) 17:03:07 -48:44:40, with a rotation angle of 76.5$^\circ$. This placed
the target and a bright reference star (K$_s$=9.5) in the lower-left quadrant (Q1). Photometric data were extracted using the ULTRACAM data reduction software tools\footnote{see also \href{http://deneb.astro.warwick.ac.uk/phsaap/ultracam/}{http://deneb.astro.warwick.ac.uk/phsaap/ultracam/}} \citep{dhilon2007}. Parameters for the extraction were derived from the bright reference star and position, to which the position of the target was linked in each exposure. To account for seeing effects, the ratio between the source and the reference star count rate was used. The time of each frame was then put in the Dynamical barycentric time system.

\section{Analysis and Results} \label{sec:style}

\begin{figure*}
 \centering
 
 \includegraphics[width=0.8\textwidth   ]{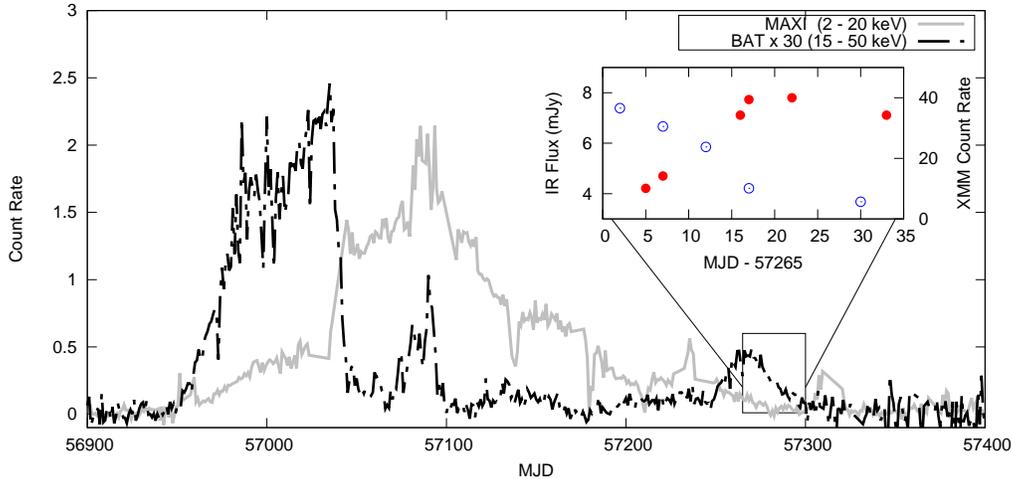}
       	\caption{ Long term light curve measured by MAXI (grey solid curve) and Swift BAT (black line-dotted curve). The inset shows the evolution of the IR flux (red full points) and the 3-10 keV X-ray count rate measured by XMM (blue empty points) as a function of time during our campaign.  The reference date for the light curve in the inset is 31/08/2015.
      	}
       	\label{fig:curve}
        \end{figure*}

The spectral analysis of the X-ray observation \citep[reported by][]{demarco2017} showed that the source had already reached the hard state during our campaign. The average IR magnitude and X-ray count rate for each epoch are plotted in Fig. \ref{fig:curve} , showing that we caught the source in its typical O-IR re-brightening \citep{dincer2012,corbel2013}, with a peak on the 17th of September.

%We analysed the long term behaviour of the source during our campaign. First we measured the average IR magnitude and X-ray count rate for each epoch (see bottom panel of  Fig. \ref{fig:frequencies}).  The spectral analysis of the X-ray observation showed that the source was in the hard state \citep[see][]{demarco2017}. While the X-ray emission is seen to gradually decrease (a factor of $\approx$ 10 in  $\approx$30 days) as function of time, the IR magnitude decreases quite rapidly reaching a peak value of 12.3 on the 17/09, starting then to slowly increase. These observations are consistent with the general phenomenological picture of  GX 339-4 during this phase of the outburst \citep{dincer2012,corbel2013} .

\begin{figure}
 \centering
 \includegraphics[width=\columnwidth]{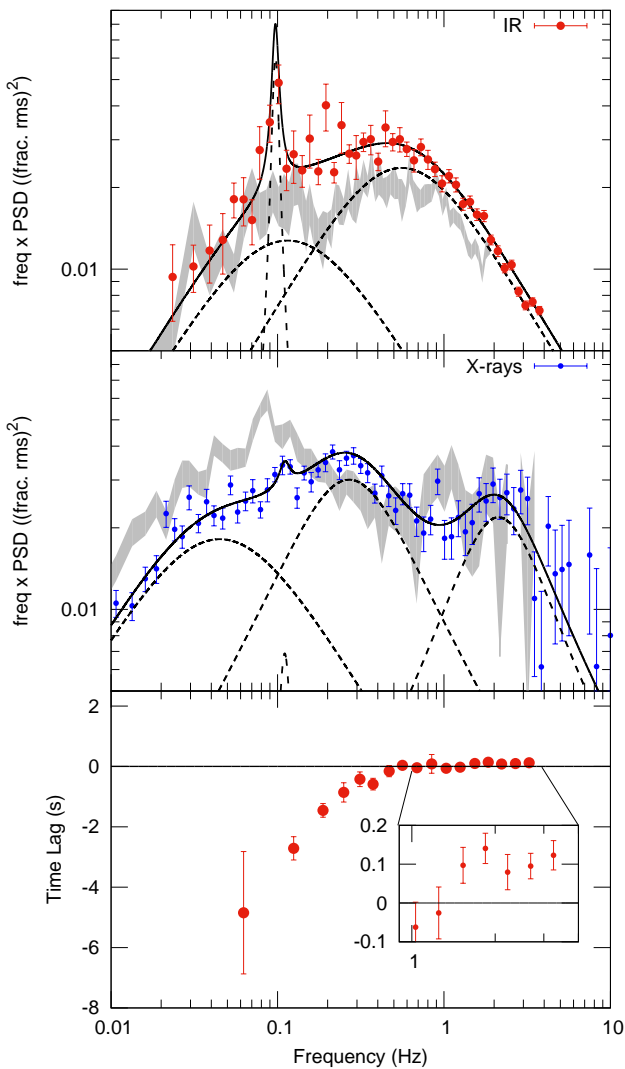}
       	\caption{ Results from the Fourier analysis of the strictly simultaneous observation on the 17th of September 2015. {\it Top panels}: PSDs computed with 1024 (IR) and 16384 (X-ray) bins per segments. The dotted lines indicate the Lorentzian components used to model the PSDs.  The grey curves represent the PSDs measured on the last epochs of the campaign, namely on the 3rd of October (X-rays) and 30th of September (IR), for comparison. {\it Bottom panel}: frequency dependent time-lags, computed with 128 bins per segment. Positive lags imply IR lagging the X-rays.   Spectral coherence
      	}
       	\label{fig:psd}
        \end{figure}

In order to quantify the variability in the X-ray and IR bands, we computed the Fast Fourier Transform of both light curves for all the observations. For the X-rays we chose 65536 bins per segment; for IR instead,  we used 256 bins per segment for the first epoch and 1024 for remaining ones, due to the structure of the data. We then calculated the PSDs, adopting a fractional squared rms normalization \citep{belloni1990}. In Figure \ref{fig:psd} (upper and middle panel) we plot the X-ray and IR PSDs from the 17th of September. All the X-ray and IR PSDs revealed the broad-band noise usually observed in the hard(-intermediate) state of BHXRBs \citep{belloni2005,dgk07}. In order to measure any possible frequency-dependent lag between the two bands, we calculated the time lags between the two bands for the only epoch where there was exact simultaneity, the 17th of September{, following the recepies described in \citet{uttley2014}.} We report the time lag in the lower panel of Fig. \ref{fig:psd}, where positive lags imply IR lagging the X-rays. At low frequencies the IR leads the X-rays by up to several seconds, while at high frequencies the IR lags the X-rays by $\sim$ 0.1 seconds. {In the frequency ranges where the lags are detected the intrinsic coherence was found to be $\approx$ 0.2 (Vincentelli et al., in prep.)}.

Following \citet{belloni2002,belloni2005} we modelled all the PSDs with a number of Lorentzian functions ($ PSD(\nu)=\frac{r^2 ~ \Delta}{2\pi} / [(\nu-\nu_0)^2+(\Delta/2)^2]$). Three broad components were needed to approximate the X-ray broad-band noise
%. We defined them as a low, intermediate and high frequency,
and we will refer to them hereafter respectively as $\nu_{X_{LF}}$,$\nu_{X_{IF}}$ and $\nu_{X_{HF}}$. One X-ray PSD (from the September 17th epoch) also showed  marginal evidence for a narrower feature at $0.11\pm 0.01$ Hz, which we identify as a type-C QPO \citep{casella2005,motta2011}. Two broad components (a low and a high frequency one, $\nu_{IR_{LF}}$ and $\nu_{IR_{HF}}$) were instead found to be sufficient to approximate the IR broad-band noise. However, in all the IR PSDs but the last one, an additional Lorentzian component (hereafter $\nu_{IR_{QPO}}$) was needed to account for the presence of a narrow feature. We identify this narrow component as a type-C QPO, as its frequency on the 17th of September is consistent with the one of the simultaneous X-ray type-C QPO.

\begin{figure*}

 \centering
 \includegraphics[width=0.7\textwidth]{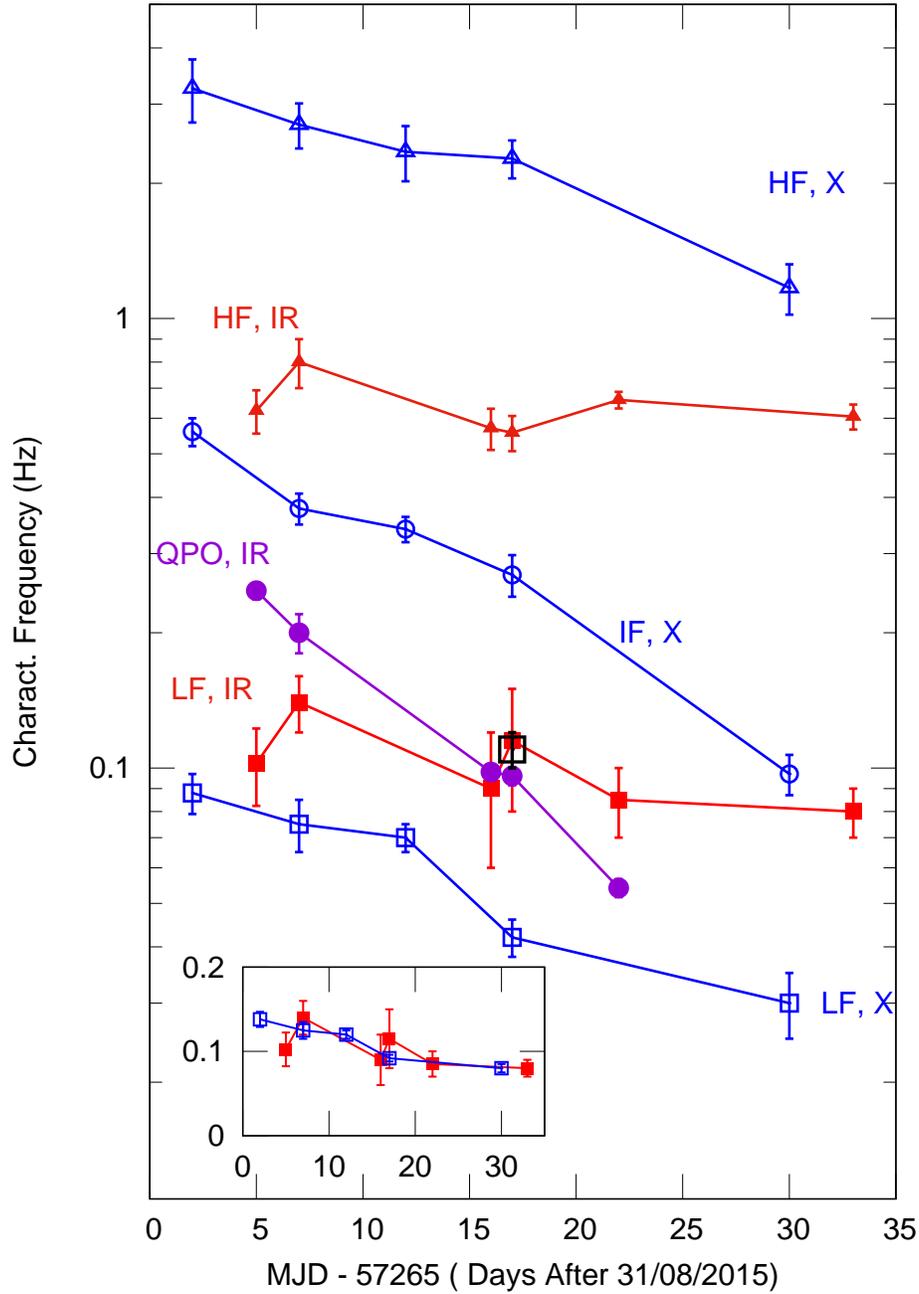}
       	\caption{ Evolution of the X-ray (blue) and IR (red) characteristic frequencies of the different Lorentzian components as a function of time. The low-frequency (LF) components are indicated with squares, while the high-frequency (HF) components with triangles. The intermediate-frequency X-ray component and the IR QPOs are indicated with circles, while the black square corresponds to the frequency of the marginally significant X-ray QPO. The inset shows a zoom on the two LF components in {\it linear} scale. We shifted $\nu_{X_{LF}}$ by +0.05 Hz, to show how the two frequencies are consistent with having a constant offset between each other.  
      	}
       	\label{fig:frequencies}
        \end{figure*}

In Fig \ref{fig:frequencies} we show the time evolution of the characteristic frequencies ($\nu_{char}=\sqrt[]{\nu_0^2+(\Delta/2)^2}$) of all the components. Linear fits as a function of time ($\nu = -a ~  t + b$) are reported  in Table \ref{tab:tab}. %give the following slopes (in $Hz/ day$): $a_{X_{LF}} = 0.0021 \pm 0.0005$, $a_{X_{IF}} = 0.014 \pm 0.0014$, $a_{X_{HF}} = 0.07 \pm 0.02$, $a_{IR_{LF}} = 0.0015 \pm 0.001$, $a_{IR_{qpo}} = 0.0105 \pm 0.0009$, and $a_{IR_{HF}} = 0.0016 \pm 0.004$ (uncertainties at 90$\%$ significance level).
We note that the slope of the high-frequency IR component is consistent with zero:  we  therefore also fitted this component with a constant, obtaining $\nu_{IR_{HF}}= 0.63 \pm 0.03$ Hz.  We also performed linear fits ($\nu_1 = a ~\nu_2 + b$) to quantify possible correlations between the low and high frequency components (see Table \ref{tab:tab}). % We find that $\nu_{X_{HF}} = (39 \pm 4) \times \nu_{X_{LF}}$ Hz, $\nu_{X_{IF}} = (6.7 \pm 0.7) \times \nu_{X_{LF}} - (0.1 \pm 0.03)$ Hz, and $\nu_{IR_{HF}} = \nu_{IR_{LF}} + (0.54 \pm0.03)$ Hz.
We could not perform direct fits to quantify possible correlations between X-ray and IR frequencies, as most of the measurements are not simultaneous. However, by linking the fits of their individual time evolution we find that the low-frequency IR component $\nu_{IR_{LF}}$ is consistent with having a constant offset of $\sim 0.05$ Hz with respect to the X-ray low-frequency component $\nu_{X_{LF}}$ (see inset in Fig. \ref{fig:frequencies}).

%$a_{IR_{LF}} = 0.0015 \pm 0.001$, $a_{IR_{qpo}} = 0.0105 \pm 0.0009$, and $a_{IR_{HF}} = 0.0016 \pm 0.004$ (uncertainties at 90$\%$ significance level).

The large uncertainties and the small number of points do not allow us to perform useful statistical tests on these correlations. This is particularly true for $\nu_{IR_{HF}}$, which is consistent with either being constant or with having a constant offset with respect to a slowly declining $\nu_{IR_{LF}}$.
% in order to disentangle between the different scenarios
%As expected and reported in the literature for many BHXRB outburst decays \citep[see e.g.][]{kalemci2004}, the characteristic frequencies of all the X-ray PSD broad components evolve in a correlated way. They decrease by about a factor of 5 in the 30 days of the campaign, while remaining proportional to each other. The evolution of the IR type-C QPO appears to follow a very similar trend.
%, which is not surprising as the OIR and the X-ray QPOs are expected to be closely connected, with at most some differences in their harmonic content \citep[see e.g.][and references therein]{kalamkar2016}. 
%The characteristic frequencies of the IR PSD broad components, instead, follow a much shallower decline, and are clearly not proportional to the X-ray components. Namely, the low-frequency IR component is consistent with having a constant offset of $\sim 0.05$ Hz with respect to the X-ray low-frequency component, and the IR high-frequency component with having a constant offset of $\sim 0.55$ Hz with respect to the IR low-frequency component. We note that the data are also consistent with the IR high-frequency component remaining constant throughout the 1-month campaign, but the small number of points does not allow us to perform useful statistical tests on these correlations. We can instead securely exclude a simple proportionality between the IR and the X-ray high-frequency components, as it is apparent from Figure \ref{fig:frequencies}.

\begin{table*}[t]
  \centering
  \begin{tabular}{ccccccc}

\\
\hline

\textbf{Frequency vs Time} & & & & & & 
\\

Band & $a_{LF}  $  & $b_{LF}$  &   $a_{IF} $ &  $b_{IF}$  & $a_{HF} $ &  $b_{HF}$   \\
\\
 & ( $10^{-2} Hz / days$ ) & (Hz) & ( $10^{-2} Hz / days$ ) & (Hz)  & ( $10^{-2} Hz / days$ ) & (Hz) 
\\
\\
  \hline

\\
 IR  & 0.15 $\pm $  0.1 & 0.08 $\pm $ 0.05 &   1.05 $\pm $ 0.09   & 0.2 $\pm $ 0.1  & 1.6 $\pm $ 4  &  0.4 $\pm $ 0.3 \\
X-rays  &0.21 $\pm $ 0.05 &0.05 $\pm $ 0.04 &  1.39 $\pm $ 0.14  &  0.3 $\pm $  0.2 & 7 $\pm $ 2 &1.9 $\pm $ 1.4     \\
   \\
   \hline
\textbf{Frequency vs Frequency} & & &

\\

Bands & $a$  & $b$  & 
\\
 &  & (Hz)  & 
 \\
\hline

\\

$\nu_{X_{HF}}$ vs $\nu_{X_{LF}}$ & 32 $\pm $9 & 0.36 $\pm $ 0.44  & & & &
\\
$\nu_{IR_{HF}}$ vs $\nu_{IR_{LF}}$ & 0 $\pm $ 2 & 0.6 $\pm $ 0.2 &  & & &
\\
\\
\hline

\end{tabular}
 
\caption{Fits of characteristic frequencies trends. The top table reports the fit of three components as a function of time ($\nu = -a ~ t + b$). Where $t$ is time in days starting from MJD 57265 In the bottom one  we show the results of the fit linking the evolution of different components within one single electromagnetic band. In particular we focused on the trend between the low and the high frequency component, i.e.  $\nu_{HF} = a ~ \nu_{LF} + b$. $\chi^2$ test revealed that the fits were all acceptable within a 95\% confidence interval. All errors are reported with a 90\% significance level.  }
\label{tab:tab}
\end{table*}

\section{Discussion}

We have discovered a peculiar evolution of the characteristic frequencies of the IR emission from the BHT GX 339--4, during its 2015 outburst decay. Namely, while the X-ray PSD shows the expected behaviour, with all its characteristic frequencies 
%of the broad-band noise
remaining proportional to each other (i.e. with a constant ratio) while 
decreasing by about a factor of 5 in about 30 days, the IR PSD reveals a relatively stable broad-band noise, with its characteristic frequencies decaying only very slowly. The IR low-frequency component is consistent with following closely the X-ray low-frequency one, with a constant offset of $\sim 0.05$ Hz (see inset in Fig \ref{fig:frequencies}). The type-C QPO evolves more rapidly, with a decaying rate relatively similar to that of the X-ray intermediate-frequency component. What is surprising is the slow decay of the IR high-frequency component, which appears clearly disconnected from the X-ray high-frequency component. Instead, it is consistent with having a constant offset of $\sim 0.54$ Hz with respect to the IR low-frequency component, as well as with being { approximately unchanging at} $\sim 0.63$ Hz throughout the campaign.

The detection of such a distinctive evolution of the IR PSD holds a strong interpretative potential, as it provides constraints to the main models discussed in the literature for the IR emission from BHTs: reprocessing from the outer disk, synchrotron emission from a magnetized inflow, and synchrotron emission from a jet. An IR low-frequency component correlated with the X-ray low-frequency one is perhaps naturally accounted for in all scenarios. The substantial stability of the IR high-frequency component is instead more constraining. The statistics of the data do not allow us to conclude whether this component is decaying very slowly or constant, but in either case it is clear that its evolution is disconnected from that of the X-ray high-frequency component. At the same time, the time lags show that the variability in the two bands is correlated over a broad range of frequencies, including those above the IR high-frequency break (Fig. \ref{fig:psd}), which implies the two signals are not independent. 

We note that, so far, all the reported fast-photometry O-IR observations of GX 339--4 have revealed a high-frequency component at similar frequencies as reported here, although never in the same outburst \citep{casella2010,gandhi2010,kalamkar2016,vincentelli2018}. This might suggest that this frequency is indeed constant, but further observations, especially longer monitoring campaigns, will be needed to confirm this.

In the following, we discuss the implications of these results in the context of possible physical scenarios.
% An IR low-frequency break correlated with the X-ray low-frequency break is perhaps naturally accounted for in both scenarios, as it is the fact that the IR variability - at these low frequencies - appears to anticipate the X-ray variability \citep{veledina2017,malzac2018}.

%The connection between the two signals is also apparent from the behaviour of the IR QPO, which follows rather closely the behaviour of the X-ray PSD components.

\subsection{Outer disk}

The observed stability of the IR high-frequency component could be easily explained if the observed IR variability were dominated by reprocessing of the X-ray emission from the outer disk. The maximum IR frequency would be then related to the disk response function, which is not expected to change with time. 

%\noindent {\it \underline{Lags}:} We note here, h
However the lags we measure are too small to be consistent with the travel time from the X-ray emitting inner regions of the flow to the IR-emitting outer regions of the disk (or even negative, see Fig. \ref{fig:psd}, lower panel). Additionally, the same 0.1-second delay has been already reported a few times for this source, and in those cases a thermal origin \citep{gandhi2008,casella2010,gandhi2017} had been robustly excluded. Thus, we conclude that the reprocessing scenario can be securely discarded.%, and we refer the reader to the discussion in the above works.

\subsection{Hot Inflow}

We consider here the hypothesis that the whole IR variable emission comes from a magnetized inflow, and that the evolution of the X-ray PSD is interpreted in terms of interference between two Comptonization continua  \citep{veledina2013,veledina2016}. 
%In this case, the observed IR variable emission would correspond to the seeds of one of the two Comptonization continua contributing to the X-ray emission, and would represent a better proxy for the mass accretion rate than the X-ray emission. 
The observed IR variable emission would then correspond to the seed synchrotron photons emitted before the Comptonization takes place in the hot inflow, and therefore can be interpreted as a proxy for the mass accretion rate. A slow decay of all the IR characteristic frequencies could then be associated with a slowly evolving geometry of the accretion inflow \citep{ingram2012,ingram2016}. 
{To maintain a constant O-IR high-frequency component within this scenario would require that the magnetized inflow extends beyond $\approx$ 50 Schwarzschild radii (or 100 gravitational radii, $R_G= G M / c^2$) during the entire  re-brightening phase. Most of the IR flux as well as variability in the magnetised inflow model originate within a region of this size \citep[][]{veledina2013}. %, i.e. increasing, however, the emitted intensity at  the O-IR wavelengths.} 
%Quantitative measurements of the actual size of the hot-inflow are still a matter of debate, mainly due to strong discrepancies found between different methods also within the same observations \citep[See e.g. ][and references therein]{Wang-Ji2018,wang2019,Mahmoud2019}. 
On the other hand, we clearly see a rise in IR flux as the source evolves through our hard state observations (see Fig. \ref{fig:curve}). Therefore, under this same scenario, O-IR flux changes cannot be attributed solely to a change in the size of the hot-inflow \citep[see e.g.][]{poutanen2014}.}

{It is also relevant that in the hard state of GX~339-4 and other BHXRBs, the soft X-ray blackbody-like emission thought to originate from the geometrically thin, optically thick accretion disk is seen to vary significantly and also lead the correlated X-ray power-law variations, over the time-scale range covered by the low and high frequency-breaks observed here \citep{wilkinson2009,uttley2011,demarco2015,demarco2017}. For geometrically thin disks, such relatively short (seconds) time-scales of large-amplitude (tens of per cent) disk continuum variability, presumably generated on the local viscous time-scale, would be difficult to explain if the disk emission originates at $>100$~$R_G$, as would be implied by the hot inflow model as presented in \citet{veledina2013}. On the other hand, on even shorter variability time-scales ($> 1$~Hz), blackbody `reverberation lags' are seen, some of which, if simply interpreted as light-travel times, could be consistent with such large radii \citep{demarco2017}, but interpretation of such lags is complex, as their combination with the continuum lags is not simply additive (e.g. see \citealt{mastroserio2019} for the case of Fe K reverberation plus continuum lags in BHXRBs).}

%New physical constraints from self-consistent spectral-timing models \citep[e.g.][]{Mahmoud2019,mastroserio2019} ought to improve significantly the estimates of the truncation radius, testing the described scenario. } % radii substantially larger than the 10-50 Schwarzschild radii invoked in the present version of the model \citep{veledina2016}, and would likely require some substantial modification. Further, it appears also to be  larger than X-ray estimates of the coronal region \citep{kara2019}.}

%However, a constant IR high-frequency component
%, \textbf{with the X-rays evolving in time}, 
%would be instead difficult to explain within this scenario, \textbf{unless the magnetised inflow extends outwards to radii unrealistically larger than the IR-emitting region \citep[][]{veledina2016,kara2019}}.
%as the geometry of the inflow is bound to evolve in order to explain the evolution of all the other frequencies.% On the other hand, however, we notice that the in this case we also expect $\nu_{IR_{LF}}$ to be lower than $\nu_{X_{LF}}$, as opposite to what is observed.

\noindent {\it \underline{Lags}:} {The negative lags measured at low frequencies are naturally expected from the hot-inflow model \citep{veledina2011,veledina2017}. On the other hand, however,} the measured positive IR delay at high frequencies seems difficult to reconcile with this scenario, as it would imply that the Comptonized signal {\it leads} its seeds. We note that modelling of this scenario has been done only for a few observations {during the intermediate state} \citep{veledina2017,veledina2018}. Therefore, new dedicated simulations are needed to test {the model also in the hard state.}%this hypothesis. 

%In particular, while the lower one is related to the truncation radius of the accretion disk, the X-ray high frequency break is associated to the inner radius of the hot inflow. This therefore would indicate that the mass accretion rate plays a key role for the low frequency components, which external boundary of the hot inflow. On the other hand, detailed modelling of interference has been done only for a few cases, using exclusively X-ray observation. Therefore the effect of interference between different components along the outburst is still not clear....

\subsection{Compact Jet - Internal shocks}

In this scenario, the whole IR variable emission is interpreted in terms of synchrotron emission from a jet, in which shells with different velocity collide and dissipate their differential energy \citep{jamil2010,malzac2014,malzac2018}. The shell Lorentz factor PSD is assumed to be identical to the X-ray PSD, which is assumed to be a proxy for the accretion rate. Fast fluctuations in the shell speed will cause the shells to collide early in the jet, dissipating and then radiating away very soon their (differential) kinetic energy. Shorter- (longer-) wavelength emission, coming from inner (outer) regions in the jet, will present faster (slower) variability. At any given wavelength, the dissipation and radiation time scales in the shock sets an upper limit on the frequency of variability that can be observed, quenching faster variability  \citep{malzac2018}. Frequency damping depends on several physical parameters, including the average Lorentz factor, the jet inclination, the average accretion rate, and the properties of the injected variability. {Thus, the observed stability of this frequency during the outburst decay, while the X-ray flux and X-ray characteristic frequencies decay \citep{belloni2005,dincer2012,demarco2017}, provides strong quantitative constraints on this scenario.}
%Therefore, as we observe this frequency to remain stable during the outburst decay, while the X-ray flux and X-ray characteristic frequencies decay \citep{belloni2005,dincer2012,demarco2017}, this result will help placing further quantitative constraints.

%a more detailed modelling will provide more quantitative constraints based on this result.% a more detailed modelling will permit to place quantitative constrains from this result.  %this result will permit to place quantitative constraints to the model.

\noindent {\it \underline{Lags}:} The IR 0.1 second lag observed at high frequencies is easily explained in terms of travel time of the fluctuations from the inflow to the jet \citep{malzac2014,malzac2018}. {\citet{malzac2018} have shown that also the negative lags observed at low frequencies can be accounted for by the model, due to a differential response of the shocks together with Doppler boosting modulation. Also in this case, dedicated simulations are in order to quantify this scenario for this dataset.} %accounted for by the model, depending on the choice of physical parameters.

%\textbf{Questo paragrafo si puo' togliere a questo punto, giusto?}It is interesting to notice that the  0.1 s O-IR lag was always detected  at high frequencies, \textit{always} in the same frequency range of a break in the O-IR PSD (\citep{gandhi2011,vincentelli2018,malzac2018}. This therefore suggests the presence of a damping mechanism which does not allow short timescales variability to be seen in the jet emission.

%A jet origin for (at least) the fast IR variability is supported by the detection of a $\sim 0.1$ s IR lag at frequencies above 1 Hz. This is in fact the same delay already reported in the literature, for which a jet origin has been - at least in some cases - securely identified \citep{gandhi2008,casella2010,kalamkar2016,gandhi2017}.

{\subsection{Disk-Jet connection - Launching radius}}

Here we focus on the possibility that the observed IR variable emission originates in a jet, but that the quenching of the IR high-frequency variability {happens at the jet-launching site, i.e. not all the variability in the inflow is transferred into the jet.}
%happens "before" the fluctuations are actually injected in the jet.
This could happen if a  \citet{blandford&payne} launching mechanism is at play \citep[as suggested by radio imaging of the jet in M87 and 3C84,][see however \citealp{liska2018}]{doeleman,giovannini} . In this case, the jet would tap matter (and accretion rate fluctuations) from a range of radii, between the innermost stable orbit and some radius $R_{launch}$ \citep{spruit2010}. Assuming that each observed X-ray Fourier frequency is associated with the viscous time scale at a given radius \citep{dgk07}, and that variability originated at any radius propagates inward through the inflow \citep{uttley2001,uttley2014}, this implies that variability (much) faster than the viscous timescale at $R_{launch}$ will contribute (much) less to the overall fluctuations transferred into the jet. %This could be applied in the case in which the two IR components evolve together  or in the case they are independent from each other

%It is reasonable to assume that each X-ray Fourier frequency is associated with the viscous time scale at a given radius \citep{dgk07}. and that the jet is launched within a range of radii \citep{spruit2010}. In this case, variability faster than the viscous timescales at these radii would originate in the inflow only {\it after} the launching region, thus would not go into the jet.

It is useful here to recall that the high-frequency component in the X-ray PSD shows a saturation at a few Hz in several BHXRBs  \citep{churazov2001,belloni2005,ingram2012}. Such a timescale is consistent with the viscous timescale at the last stable orbit ($R_{ISCO}$) of a hot inflow around a black-hole of $\approx 10 M_\odot$. For GX 339--4, this (assumed) viscous frequency at $R_{ISCO}$ is $\approx 2-5$ Hz \citep{dgk07}, i.e. $\approx 4-10$ times the value we measure for the high-frequency IR component. If the latter corresponds to the viscous frequency $\nu_{visc}$ at the launching radius, as $\nu_{visc}  \propto  R^{-3/2}$, it follows that the size of the launching region would be $R_{launch}=(4-10)^{2/3} R_{ISCO}$, i.e. of the order of $\sim 10  R_G$. We caution the reader that this estimate is very rough, as it depends on many assumptions and on unknown parameters, such as the thickness and viscosity of the hot flow, and serves only as a figure of merit to test the scenario.
We note that, {in past outbursts from this source,} the IR flux from GX 339--4 {has been observed to drop by 3 magnitudes in just a few days - as the source leaves the hard state -  when the X-ray low-frequency component is at a frequency $\approx$\,0.55 Hz  \citep{belloni2005,homan2005}. This frequency is very close to the high-frequency IR component we measure,
%. While it is not directly clear whether this is quantitatively expected by the magnetised inflow scenario, it
which is consistent with a scenario in which the inferred inflow becomes comparable in size to the jet launching region, causing the jet to quench.}

%, but it appears interestingly close to the measured 5.5 $R_G$ jet launching region in the supermassive black hole in M87, obtained through radio imaging of the parabolic jet structure from $10^4$-$10^6$ $R_G$ scales down to $\approx 10 R_G$ from the central black-hole \citep{hada2011,doeleman}.

\noindent {\it \underline{Lags}:} Also in this case, the IR 0.1 second lag observed at high frequencies can be intuitively explained in terms of travel time of the fluctuations from the inflow to the jet. The low-frequency lags are instead more difficult to interpret, as {this specific disk-jet} scenario focuses on the high-frequency variability. Any prediction on the low-frequency behaviour would be necessarily model dependent, and is beyond the scope of this work.

\section{conclusions}

We have analysed data from a quasi-simultaneous, multi-epoch X-ray/IR fast-photometry campaign of GX 339--4 during its 2015 outburst decay. We studied the evolution of the Fourier power density spectra in the two wavelengths. Our main result is the discovery of the stability of the characteristic frequencies of the IR variability, in particular of the high-frequency component, which is consistent with remaining constant at $\sim$0.63 Hz throughout the campaign. We discuss possible physical interpretations of this result, and conclude that the most plausible explanation is in terms of variability from the X-ray emitting inflow being transferred into an IR emitting jet{, although all models will need to be tested quantitatively against this result. Future multi-wavelength campaigns, with multiple simultaneous bands in the optical-IR range, will help to disentangle between the scenarios.}

\section{acknowledgments}
 BDM acknowledges support from the European Union's Horizon 2020 research and innovation programme under the Marie Sk{\l}odowska-Curie grant agreement No 798726. JM and POP acknowledges financial support from PNHE in France and  JM  acknowledges support from the OCEVU Labex (ANR-11-LABX-0060) and the A*MIDEX project (ANR-11-IDEX-0001-02) funded by the ”Investissements d’Avenir” French government program managed by the ANR. PG acknowledges support from STFC (ST/R000506/1).

\bibliographystyle{mnras}
\bibliography{bib} % if your bibtex file is called example.bib



\end{document}